\documentclass[published]{JHEP3} 

\JHEP{00(2007)000}

\JHEPspecialurl{http://jhep.sissa.it/JOURNAL/JHEP3.tar.gz}

\usepackage{epsfig,multicol,bbm}

\newcommand\fverb{\setbox\fverbbox=\hbox\bgroup\verb}
\newcommand\fverbdo{\egroup\medskip\noindent%
			\fbox{\unhbox\fverbbox}\ }
\newcommand\fverbit{\egroup\item[\fbox{\unhbox\fverbbox}]}
\newbox\fverbbox

\def\a{\alpha}
\def\b{\beta}
\def\g{\gamma}
\def\d{\delta}
\def\D{\Delta}

\def\t{\theta}

\def\l{\lambda}
\def\m{\mu}

\def\be{\begin{equation}}
\def\ee{\end{equation}}
\def\ba{\begin{eqnarray}}
\def\ea{\end{eqnarray}}

\newcommand{\rpar}{\stackrel{\leftarrow}{\partial}}
\newcommand{\lpar}{\stackrel{\rightarrow}{\partial}}

\newcommand{\no}{\nonumber}

\title{Nambu-like odd brackets on supermanifolds}

\author{Dmitrij V. Soroka and Vyacheslav A. Soroka\\
        Kharkov Institute of Physics and Technology\\
        1, Akademicheskaya St., 61108 Kharkov, Ukraine\\
        E-mail: \email{dsoroka@kipt.kharkov.ua}, \email{vsoroka@kipt.kharkov.ua}}

\received{}
\accepted{}

\preprint{\hepth{9912999}}      

\abstract{The Grassmann-odd Nambu-like brackets corresponding to an arbitrary 
Lie superalgebra and realized on the supermanifolds are proposed.}


\dedicated{Dedicated to Valentina Erofeyevna Korol' and Nikita Dmitriyevich 
Soroka.\\}

\begin{document} 


\section{Introduction}

In the paper \cite{s1} the Grassmann-odd linear Poisson bracket corresponding 
to the $SO(3)$ group and built up only of the Grassmann variables has been
introduced. It was found for this bracket at once three Grassmann-odd 
nilpotent Batalin-Vilkovisky type $\D$-like differential operators 
\cite{bv1,bv2} of the first, second and third orders with respect to the 
Grassmann derivatives. Then \cite{ss1} the Nambu \cite{n} Grassmann-odd
bracket on the Grassmann algebra was constructed with use of the third order
$\D$-operator.

In \cite{s2,s3,ss2} the linear Grassmann-odd Poisson bracket corresponding to 
an arbitrary Lie group and realized solely on the Grassmann 
variables has been proposed. Again it was found
for this bracket three Grassmann-odd nilpotent $\D$-like differential 
operators of the first, second and third orders with respect to the Grassmann 
derivatives. Later \cite{ss3} with the help of the third order $\D$-operator 
the Nambu-like Grassmann-odd bracket on the Grassmann algebra was developed.

The purpose of the present paper is to extend this construction onto the 
arbitrary Lie supergroup realized on the supermanifolds. In 
particular, the linear Grassmann-odd Poisson bracket corresponding to 
the arbitrary Lie supergroup is constructed on the supermanifolds.
Then for this bracket six Grassmann-odd nilpotent $\D$-like differential 
operators of the first, second and third orders with respect to the derivatives
are obtained. At last, for every Lie supergroup by means of two third order 
$\D$-operators two different Nambu-like Grassmann-odd brackets are built up on 
the supermanifolds.

\section{Linear odd Poisson bracket on supermanifolds}

The operator $\Pi$ of the Grassmann parity inversion (see, e.g., \cite{aksz})
acts on the coordinates $z_\a$ of the co-adjoint representation $G^*$ of the
Lie superalgebra $G$ in the following way:
\ba
\Pi z_\a=c_\a,\qquad p(c_\a)=p(z_\a)+1\pmod2,\no
\ea
where $c_\a$ are coordinates in $\Pi G^*$ and $p(z_\a)$ is a Grassmann parity 
of the value $z_\a$.

The linear Grassmann-odd Poisson bracket corresponding to the arbitrary 
Lie supergroup and realized on the supermanifolds has the following 
form:
\ba\label{2.1}
\{A,B\}_1=A\rpar_{c_\a}{f_{\a\b}}^\g c_\g\lpar_{c_\b}B,
\ea
where ${f_{\a\b}}^\g$ are structure constants entering in the permutation 
relations 
\ba
G_\a G_\b-(-1)^{p(z_\a)p(z_\b)}G_\b G_\a={f_{\a\b}}^\g G_\g\no
\ea
for the generators $G_\a$ of the Lie superalgebra $G$,
$\rpar$ and $\lpar$ are the right and left derivatives,  
$\partial_{c_\a} \equiv {\partial \over {\partial c_\a}}$ and $A, B$ are 
functions of $c_\a$. The structure constants ${f_{\a\b}}^\g$ have a Grassmann 
parity
\ba
p({f_{\a\b}}^\g)=p(z_\a)+p(z_\b)+p(z_\g)=0\pmod2,\no
\ea
following symmetry properties:
\ba\label{2.2}
{f_{\a\b}}^\g=-(-1)^{p(z_\a)p(z_\b)}{f_{\b\a}}^\g
\ea
and obey the Jacobi identity
\ba\label{2.3}
\sum_{(\a\b\g)}(-1)^{p(z_\a)p(z_\g)}{f_{\a\l}}^\m{f_{\b\g}}^\l,
\ea
where the symbol $(\a\b\g)$ means a cyclic permutation of the quantities
$\a, \b$ and $\g$.

The bracket (\ref{2.1}) possesses at once six Grassmann-odd nilpotent 
Batalin-Vilkovisky type \cite{bv1,bv2} $\D$-like differential 
operators of the first, second and third orders with respect to the  
derivatives
\ba\label{2.4}
\D_{+1}={1\over2}(-1)^{p(c^\a)}c^\a c^\b{f_{\b\a}}^\g\partial_{c^\g},
\ea
\ba\label{2.5}
{\tilde\D}_{+1}={1\over2}(-1)^{p(c^\a)}c^\a c^\b{f_{\a\b}}^\g\partial_{c^\g},
\ea
\ba\label{2.6}
\D_{-1}={1\over2}(-1)^{p(c_\a)}{f_{\a\b}}^\g c_\g\partial_{c_\b}
\partial_{c_\a},
\ea
\ba
{\tilde\D}_{-1}={1\over2}(-1)^{p(c_\b)}{f_{\a\b}}^\g c_\g\partial_{c_\a}
\partial_{c_\b},\no
\ea
\ba\label{2.7}
\D_{-3}={1\over3!}(-1)^{p(c_\b)+1}f_{\g\b\a}\partial_{c_\a}\partial_{c_\b}
\partial_{c_\g},
\ea
\ba\label{2.8}
{\tilde\D}_{-3}={1\over3!}(-1)^{p(c_\b)+1}f_{\a\b\g}\partial_{c_\a}
\partial_{c_\b}\partial_{c_\g}.
\ea

In the relations (\ref{2.4}) and (\ref{2.5}) $c^\a$ are coordinates in the 
adjoint representation of $G$ which for an arbitrary semi-simple Lie 
superalgebra connected with co-adjoint representation with the help of the 
metric $g^{\b\a}$
\ba
c^\a=c_\b g^{\b\a}\no
\ea
that is inverse
\ba
g_{\a\b}g^{\b\g}=\d_\a^\g\no
\ea
to the Cartan-Killing metric tensor
\ba
g_{\a\b}=(-1)^{p(z_\l)}{f_{\a\g}}^\l{f_{\b\l}}^\g.\no
\ea
The tensor with the low indices
\ba
f_{\a\b\g}={f_{\a\b}}^\l g_{\l\g}\no
\ea
in the relations (\ref{2.7}) and (\ref{2.8}), due to the equations (\ref{2.2})
and (\ref{2.3}), has the following symmetry properties:
\ba
f_{\a\b\g}=-(-1)^{p(z_\a)p(z_\b)}f_{\b\a\g}=-(-1)^{p(z_\b)p(z_\g)}f_{\a\g\b}.
\no
\ea

The operator $\D_{+1}$ (\ref{2.4}) is proportional to the second terms in the 
BRST-like nilpotent charges
\ba
c^\a G_\a-\D_{+1},\no
\ea
\ba
(-1)^{p(c^\a)+1}c^\a G_\a-\D_{+1}.\no
\ea
The operator $\D_{-1}$ (\ref{2.6}) related to the divergence of the vector 
field $\{c_\a,A\}_1$
\ba
\D_{-1}A={1\over2}(-1)^{p(c_\a)}\partial_{c_\a}\{c_\a,A\}_1\no
\ea
is proportional to the true Batalin-Vilkovisky $\D$-operator \cite{bv1,bv2} 
for the linear odd Poisson bracket (\ref{2.1}) and to the second terms in the 
following nilpotent charges:
\ba
G_\a\partial_{c_\a}-\D_{-1},\no
\ea
\ba
(-1)^{p(c_\a)+1}G_\a\partial_{c_\a}-\D_{-1}.\no
\ea
The quantities $c^\a$ and $\partial_{c^\a}$ represent the operators for the 
ghosts and ghost momenta, respectively.
The operator $\D_{-1}$ (\ref{2.6}) determines the linear odd Poisson bracket 
(\ref{2.1}) as a deviation of the Leibniz rule under the usual multiplication
\begin{eqnarray}
\Delta_{-1}(A\cdot B)=(\Delta_{-1}A)\cdot B
+(-1)^{p(A)}A\cdot\Delta_{-1}B
+(-1)^{p(A)}\{A,B\}_1\ .\nonumber
\end{eqnarray}
and simultaneously satisfies the Leibniz rule with respect to the linear odd 
Poisson bracket composition
\ba
\D_{-1}(\{A,B\}_1)=\{\D_{-1}A,B\}_1+(-1)^{p(A)+1}\{A,\D_{-1}B\}_1.\no
\ea

In the present paper we show that the operators $\D_{-3}$ (\ref{2.7}) and 
${\tilde\D}_{-3}$ (\ref{2.8}) are related with the Grassmann-odd Nambu-like 
brackets \cite{n} on the supermanifolds. These brackets correspond to the 
arbitrary Lie superalgebra.

\section{Nambu-like odd brackets}

By applying the operators $\D_{-3}$ (\ref{2.7}) and 
${\tilde\D}_{-3}$ (\ref{2.8}) to the usual product of two 
quantities $A$ and $B$, we obtain
\begin{eqnarray}
\Delta_{-3}(A\cdot B)=(\Delta_{-3}A)\cdot B
+(-1)^{p(A)}A\cdot\Delta_{-3}B
+(-1)^{p(A)}\D(A,B),\no
\end{eqnarray}
\begin{eqnarray}
{\tilde\Delta}_{-3}(A\cdot B)=({\tilde\Delta}_{-3}A)\cdot B
+(-1)^{p(A)}A\cdot{\tilde\Delta}_{-3}B
+(-1)^{p(A)}{\tilde\D}(A,B),\no
\end{eqnarray}
where the values $\D(A,B)$ and ${\tilde\D}(A,B)$ are
\ba\label{3.1}
\D(A,B)&=&{1\over2}f_{\g\b\a}(-1)^{p(c_\b)+1}
\D\left(\partial_{\t_\a}A,\partial_{\t_\g}C\right)\Bigl[(-1)^{p(A)[p(c_\g)+1]}
\left(\partial_{c_\a}\partial_{c_\b}A
\right)\partial_{c_\g}B\cr&+&(-1)^{p(A)p(c_\a)}\left(\partial_{c_\a}A\right)
\partial_{c_\b}\partial_{c_\g}B
\D\left(\partial_{\t_\a}A,\partial_{\t_\g}C\right)\Bigr],
\ea
\ba\label{3.2}
{\tilde\D}(A,B)&=&{1\over2}f_{\a\b\g}(-1)^{p(c_\b)+1}
\D\left(\partial_{\t_\a}A,\partial_{\t_\g}C\right)\Bigl[(-1)^{p(A)[p(c_\g)+1]}
\left(\partial_{c_\a}\partial_{c_\b}A
\right)\partial_{c_\g}B\cr&+&(-1)^{p(A)p(c_\a)}\left(\partial_{c_\a}A\right)
\partial_{c_\b}\partial_{c_\g}B
\D\left(\partial_{\t_\a}A,\partial_{\t_\g}C\right)\Bigr].
\ea

By acting with the operators $\D_{-3}$ (\ref{2.7}) and 
${\tilde\D}_{-3}$ (\ref{2.8}) on the usual product of three quantities
$A$, $B$ and $C$, we come to the following relations:
\ba\label{3.3}
\D_{-3}(A\cdot B\cdot C)&=&\left(\D_{-3}A\right)\cdot B\cdot C+
(-1)^{p(A)}A\cdot \left(\D_{-3}B\right)\cdot C+
(-1)^{p(A)+p(B)}A\cdot B\cdot\D_{-3}C\cr&+&(-1)^{p(A)}\D(A,B)C+
(-1)^{p(A)p(B)+p(A)+p(B)}B\D(A,C)\cr&+&(-1)^{p(A)+p(B)}A\D(B,C)+
(-1)^{p(B)}\{A,B,C\}_1,
\ea
\ba\label{3.4}
{\tilde\D}_{-3}(A\cdot B\cdot C)&=&\left({\tilde\D}_{-3}A\right)\cdot B\cdot C+
(-1)^{p(A)}A\cdot \left({\tilde\D}_{-3}B\right)\cdot C+
(-1)^{p(A)+p(B)}A\cdot B\cdot{\tilde\D}_{-3}C\cr&+&(-1)^{p(A)}{\tilde\D}(A,B)C+
(-1)^{p(A)p(B)+p(A)+p(B)}B{\tilde\D}(A,C)\cr&+&(-1)^{p(A)+p(B)}A{\tilde\D}(B,C)
+(-1)^{p(B)}\{A,B,C\}_{\tilde1},
\ea
where the last terms in the right hand sides of the (\ref{3.3}) and 
(\ref{3.4}) are the Grassmann-odd Nambu-like brackets
\ba\label{3.5}
\{A,B,C\}_1=(-1)^{p(A)[p(c_\a)+1]+p(B)[p(c_\g)+1]+p(c_\b)+1}f_{\g\b\a}
\partial_{c_\a}A\partial_{c_\b}B\partial_{c_\g}C
\ea
\ba\label{3.6}
\{A,B,C\}_{\tilde1}=(-1)^{p(A)[p(c_\a)+1]+p(B)[p(c_\g)+1]+p(c_\b)+1}f_{\a\b\g}
\partial_{c_\a}A\partial_{c_\b}B\partial_{c_\g}C
\ea
on the supermanifolds with coordinates $c_\a$.

The divergences of the Nambu-like odd brackets (\ref{3.5}) and (\ref{3.6}) are 
related with the values $\D(A,B)$ (\ref{3.1}) and ${\tilde\D}(A,B)$ 
(\ref{3.2}) (see also \cite{sak})
\ba
\D(A,B)={1\over2}\partial_{c_\a}\{c_\a,A,B\}_1,\no
\ea
\ba
{\tilde\D}(A,B)={1\over2}\partial_{c_\a}\{c_\a,A,B\}_{\tilde1}.\no
\ea
The contraction of the Grassmann-odd Nambu-like bracket (\ref{3.5}) with the 
variable $c^\a$ gives the linear odd Poisson bracket (\ref{2.1})
\ba
\{A,B\}_1=-c^\a\{c_\a,A,B\}_1,\no
\ea
whereas the contraction of the odd Nambu-like bracket (\ref{3.6}) with $c^\a$ 
gives the expression
\ba
c^\a\{c_\a,A,B\}_{\tilde1}=(-1)^{p(c_\a)p(c_\b)+1}
A\rpar_{c_\a}{f_{\a\b}}^\g c_\g\lpar_{c_\b}B,\no
\ea
which is not a Poisson bracket, since it does not obey the Jacobi identity.
Note also the following relations between the operators $\D_{-3}$ (\ref{2.7}), 
${\tilde\D}_{-3}$ (\ref{2.8}) and odd Nambu-like brackets (\ref{3.6}), 
(\ref{3.5}):
\ba
\D_{-3}A={1\over3!}(-1)^{p(c_\a)}\partial_{c_\a}\partial_{c_\b}
\{c_\a,c_\b,A\}_{\tilde1},\no
\ea
\ba
{\tilde\D}_{-3}A={1\over3!}(-1)^{p(c_\a)}\partial_{c_\a}\partial_{c_\b}
\{c_\a,c_\b,A\}_1.\no
\ea

For the operator $\D_{-3}$ (\ref{2.7}) and ${\tilde\D}_{-3}$ (\ref{2.8}) there 
exists the following ``Leibniz rules'' with respect to the odd Nambu-like 
bracket compositions:
\ba
\D_{-3}(\{A,B,C\}_1)=&-&\{\D_{-3}A,B,C\}_1+(-1)^{p(A)}\{A,\D_{-3}B,C\}_1\cr
&-&(-1)^{p(A)+p(B)}\{A,B,\D_{-3}C\}_1\cr&+&(-1)^{p(A)[p(c_\a)+1]
+p(B)[p(c_\g)+1]}f_{\g\b\a}\cr&\times&\Bigl[(-1)^{p(A)+p(c_\g)}
\D\left(\partial_{c_\a}A,\partial_{c_\b}B\right)\partial_{c_\g}C\cr
&+&(-1)^{p(A)+p(B)+p(c_\a)+1}\partial_{c_\a}A\D\left(\partial_{c_\b}B,
\partial_{c_\g}C\right)\cr&+&(-1)^{p(A)p(B)+p(c_\b)+p(A)[p(c_\b)+1]
+p(B)[p(c_\a)+1]+[p(c_\a)+1][p(c_\b)+1]}\cr&\times&\partial_{c_\b}B
\D\left(\partial_{c_\a}A,\partial_{c_\g}C\right)\Bigr],\no
\ea
\ba
{\tilde\D}_{-3}(\{A,B,C\}_{\tilde1})=&-&\{{\tilde\D}_{-3}A,B,C\}_{\tilde1}
+(-1)^{p(A)}\{A,{\tilde\D}_{-3}B,C\}_{\tilde1}\cr
&-&(-1)^{p(A)+p(B)}\{A,B,{\tilde\D}_{-3}C\}_{\tilde1}\cr
&+&(-1)^{p(A)[p(c_\a)+1]
+p(B)[p(c_\g)+1]}f_{\a\b\g}\cr&\times&\Bigl[(-1)^{p(A)+p(c_\g)}
{\tilde\D}\left(\partial_{c_\a}A,\partial_{c_\b}B\right)\partial_{c_\g}C\cr
&+&(-1)^{p(A)+p(B)+p(c_\a)+1}\partial_{c_\a}A{\tilde\D}\left(\partial_{c_\b}B,
\partial_{c_\g}C\right)\cr&+&(-1)^{p(A)p(B)+p(c_\b)+p(A)[p(c_\b)+1]
+p(B)[p(c_\a)+1]+[p(c_\a)+1][p(c_\b)+1]}\cr&\times&\partial_{c_\b}B
{\tilde\D}\left(\partial_{c_\a}A,\partial_{c_\g}C\right)\Bigr],\no
\ea
\ba
\D_{-3}(\{A,B,C\}_{\tilde1})=&-&\{\D_{-3}A,B,C\}_{\tilde1}
+(-1)^{p(A)}\{A,\D_{-3}B,C\}_{\tilde1}\cr
&-&(-1)^{p(A)+p(B)}\{A,B,\D_{-3}C\}_{\tilde1}\cr&+&(-1)^{p(A)[p(c_\a)+1]
+p(B)[p(c_\g)+1]}f_{\a\b\g}\cr&\times&\Bigl[(-1)^{p(A)+p(c_\g)}
\D\left(\partial_{c_\a}A,\partial_{c_\b}B\right)\partial_{c_\g}C\cr
&+&(-1)^{p(A)+p(B)+p(c_\a)+1}\partial_{c_\a}A\D\left(\partial_{c_\b}B,
\partial_{c_\g}C\right)\cr&+&(-1)^{p(A)p(B)+p(c_\b)+p(A)[p(c_\b)+1]
+p(B)[p(c_\a)+1]+[p(c_\a)+1][p(c_\b)+1]}\cr&\times&\partial_{c_\b}B
\D\left(\partial_{c_\a}A,\partial_{c_\g}C\right)\Bigr],\no
\ea
\ba
{\tilde\D}_{-3}(\{A,B,C\}_1)=&-&\{{\tilde\D}_{-3}A,B,C\}_1
+(-1)^{p(A)}\{A,{\tilde\D}_{-3}B,C\}_1\cr
&-&(-1)^{p(A)+p(B)}\{A,B,{\tilde\D}_{-3}C\}_1\cr&+&(-1)^{p(A)[p(c_\a)+1]
+p(B)[p(c_\g)+1]}f_{\g\b\a}\cr&\times&\Bigl[(-1)^{p(A)+p(c_\g)}
{\tilde\D}\left(\partial_{c_\a}A,\partial_{c_\b}B\right)\partial_{c_\g}C\cr
&+&(-1)^{p(A)+p(B)+p(c_\a)+1}\partial_{c_\a}A{\tilde\D}\left(\partial_{c_\b}B,
\partial_{c_\g}C\right)\cr&+&(-1)^{p(A)p(B)+p(c_\b)+p(A)[p(c_\b)+1]
+p(B)[p(c_\a)+1]+[p(c_\a)+1][p(c_\b)+1]}\cr&\times&\partial_{c_\b}B
{\tilde\D}\left(\partial_{c_\a}A,\partial_{c_\g}C\right)\Bigr].\no
\ea

The Grassmann parity
\ba
p(\{A,B,C\})=p(A)+p(B)+p(C)+1\pmod2,\no
\ea
symmetry properties
\ba
\{A,B,C\}=-(-1)^{[p(A)+1][p(B)+1]}\{B,A,C\}
=-(-1)^{[p(B)+1][p(C)+1]}\{A,C,B\}\no
\ea
and Jacobi type identity 
\ba\label{3.7}
\{\{A,B,C\},D,E\}&+&(-1)^{[p(A)+1][p(B)+p(C)+p(D)+p(E)]}
\{\{B,C,D\},E,A\}\cr&+&(-1)^{[p(A)+p(B)][p(C)+p(D)+p(E)+1]}
\{\{C,D,E\},A,B\}\cr&+&(-1)^{[p(D)+p(E)][p(A)+p(B)+p(C)+1]}
\{\{D,E,A\},B,C\}\cr&+&(-1)^{[p(E)+1][p(A)+p(B)+p(C)+p(D)]}
\{\{E,A,B\},C,D\}\cr&+&(-1)^{[p(D)+p(E)][p(A)+p(B)+p(C)+1]+p(B)[p(A)+1]
+p(A)}\cr&\times&\{\{D,E,B\},A,C\}\cr&+&(-1)^{[p(D)+1][p(A)+p(B)+p(C)]
+p(D)}
\{\{D,A,B\},C,E\}\cr&+&(-1)^{[p(A)+1][p(B)+p(C)+p(D)+p(E)]+[p(D)+1]p(E)
+p(D)}\cr&\times&\{\{B,C,E\},D,A\}\cr&+&(-1)^{[p(B)+1][p(C)+p(D)+p(E)]
+p(B)}
\{\{A,C,D\},E,B\}\cr&+&(-1)^{[p(B)+1][p(C)+p(D)+p(E)+1]+[p(D)+1][p(E)+1]}
\cr&\times&\{\{A,C,E\},D,B\}=0
\ea
follow from the expressions (\ref{3.5})  and (\ref{3.6}) for the odd 
Nambu-like brackets, where either
\ba
\{A,B,C\}\equiv\{A,B,C\}_1\no
\ea
or
\ba
\{A,B,C\}\equiv\{A,B,C\}_{\tilde1}.\no
\ea

Note that the structure of (\ref{3.7}) is different from the structure of the
"fundamental identity" \cite{t}, which has four terms, and, containing ten
terms, is similar to the structure of the
"generalized Jacobi identity" \cite{appb,az1,az2} and to the
``(J2)-structure'' \cite{vv}, which is intimately related to the homotopy
algebras \cite{ss} and SH-algebras \cite{ls}.

\section{Conclusion}

Thus, by using the Grassmann-odd nilpotent differential operators $\D_{-3}$ 
(\ref{2.7}) and ${\tilde\D}_{-3}$ (\ref{2.8}) of
the third order with respect to the derivatives, we constructed two different
Grassmann-odd Nambu-like brackets (\ref{3.5}) and (\ref{3.6}), respectively, 
which correspond to 
the arbitrary Lie superalgebra and are realized on the 
supermanifolds. The main properties of
these brackets are also given.

It would be interesting to consider the dynamics based on these brackets. The
work in this direction is in progress.

\acknowledgments

We are grateful to J.A. de Azcarraga and J.D. Stasheff for the interest in
this work and for the attraction of our attention
to the papers \cite{appb,az1,az2} and \cite{vv}, respectively.
One of the authors (V.A.S.) thanks the administration of the Office of 
Associate and Federation Schemes of the Abdus Salam ICTP for the kind 
hospitality at Trieste where this work has been started.
The research of V.A.S. was partially supported by the Ukrainian National 
Academy of Science and Russian Fund of Fundamental Research, Grant No 
38/50-2008.

\end{document}